\documentclass[12pt]{article}

\usepackage{epsf}
\usepackage[dvips]{graphicx}

\begin{document}

\begin{center}
{\bfseries FINAL STATE INTERACTION IN NEUTRON DEUTERON CHARGE EXCHANGE
REACTION AT SMALL TRANSFER MOMENTUM}

\vskip 5mm

N.B.Ladygina$^\dag$

\vskip 5mm

{\small
 {\it
Laboratory of High Energies,
Joint Institute for Nuclear Research, 141980 Dubna, Russia,
}
\\
$\dag$ {\it
E-mail: ladygina@sunhe.jinr.ru
}}
\end{center}

\vskip 5mm

\begin{center}
\begin{minipage}{150mm}
\centerline{\bf Abstract}
Analysis of the $nd\to p(nn)$ reaction in a Gev-energy region is performed in the framework
based on the multiple-scattering theory for the few nucleon system.
The special kinematic condition, when momentum transfer from neutron beam to final
proton closes to zero, is considered. The possibility to extract the spin-flip term of
the elementary $np\to pn $  amplitude from nd-breakup process is investigated.
The energy dependence of the ratio
 $R=\frac{d\sigma_{nd}} {d\Omega} / \frac{d\sigma_{np}}{d\Omega}$
is obtained taking account of the final state interaction two outgoing neutrons
 in $^1 S_0$-state.
\end{minipage}
\end{center}

\vskip 10mm

\section{Introduction}

The nucleon- deuteron charge exchange reaction is the subject of the 
investigation in the set of the experiments, which are started
in VBLHE JINR at STRELA and DELTA SIGMA \cite {delta} setups and 
in COSY\cite {cosy} at
ANKE spectrometer. The experiments are performed in the special
kinematics, when transfer momentum from initial nucleon
to outgoing fast nucleon is close to zero. The goal of these
experiments is to extract the additional information about
spin dependent part of the elementary $np\to pn$ process
from nucleon- deuteron reaction. This idea was suggested by Pomeranchuk \cite {Pom}
already in 1951. Later, it was shown, that in the
plane-wave impulse approximation (PWIA) the differential cross
section and tensor analyzing power $T_{20}$ in the dp-charge exchange
reaction are actually fully determined  by the spin-dependent
part of the elementary $np\to pn$
 amplitudes \cite {alad}, \cite {cosy}.

However, under kinematical conditions,
when momentum of the emitted fast nucleon has the same direction and value
as the beam  (in the deuteron rest frame), and relative momentum
of the  two slow nucleons is  small, the final state interaction (FSI)
effects play very important role. The study of the FSI influence is the
goal of this paper. 

Here the $nd\to pnn$ reaction is considered in kinematics of the
DELTA SIGMA experiment \cite {delta}, when outgoing proton has the  same direction as 
projectile neutron and transfer momentum is close to zero.
The kinetic energy of the initial neutron varies from 0.8 up to 1.3 GeV.
The analysis has been performed in the deuteron rest frame.
The theoretical
approach is based on the Alt-Grassberger-Sandhas
formulation of  the multiple-scattering theory for the three-nucleon system.
The matrix inversion method 
has been applied for description of the two slow neutrons interaction.

\section{Theoretical formalism}

In accordance to the three-body collision theory, the amplitude of 
the neutron deuteron charge exchange reaction,
\begin{eqnarray}
n(\vec p)+d(\vec 0)\to p(\vec p_1)+n(\vec p_2)+n(\vec p_3)
\end{eqnarray}
is defined by the matrix element of the transition operator $U_{01}$
\begin{eqnarray}
\label{ampl}
{U}_{nd\to pnn} \equiv \sqrt {2} <123|[1-(1,2)-(1,3)] U_{01}|1(23)>=
\delta (\vec p -\vec p_1-\vec p_2 -\vec p_3){\cal J}.
\end{eqnarray}
As consequence  of the particle identity in initial and final states
 the permutation operators for two nucleons $(i,j)$ appear in this expression.

As it was shown in   ref.\cite {LSh}, the matrix element 
$U_{nd \to pnn}$ can be presented as

\begin{eqnarray}
\label{am}
U_{nd \to pnn}&=&\sqrt {2} <123|[1-(2,3)][1+t_{23}(E-E_1) 
g_{23} (E-E_1)]t_{12}^{sym}|1(23)>,
\end{eqnarray}
where the operator $g_{23} (E-E_1)$ is a free propagator for the
(23)-subsystem and the scattering operator $t_{23}(E-E_1)$
satisfies the Lippmann-Schwinger (LS)
equation with two-body force operator $V_{23}$ as  driving term
\begin{eqnarray}
\label{LS}
t_{23}(E-E_1) = V_{23} + V_{23} g_{23}(E-E_1) t_{23}(E-E_1) .
\end{eqnarray}
Here $E$ is the total energy of the three-nucleon system 
$E=E_1+E_2+E_3$.

Let us rewrite the matrix element (\ref{am}) indicating explicitly
the particle quantum numbers,
\begin{eqnarray}
U_{nd\to pnn}=\sqrt {2}
<\vec {p_1} m_1 \tau_1,\vec {p_2} m_2 \tau_2,\vec {p_3} m_3 \tau_3|
[1-(2,3)] \omega_{23} t^{sym}_{12} |\vec {p} m \tau ,\psi _{1 M_d 0 0} (23)>,
\nonumber
\end{eqnarray}
where $\omega_{23}=[1+t_{23}(E-E_1) g_{23} (E-E_1)]$ and
the the spin and isospin  projections denoted as
$m$ and $\tau$, respectively.  The operator $t_{12}^{sym}$ is symmetrized
NN-operator, $t_{12}^{sym}=[1-(1,2)]t_{12}$.

Under kinematical conditions, when
transfer momentum  $\vec q=\vec p -\vec p_1 $ is close to zero, one can anticipate
that the FSI in the $^1S_0$ state is prevalent at comparatively small
$p_0$-values.
In such a way we get the following expression for the amplitude of
the $nd$ charge exchange process \cite {EPJ}
\begin{eqnarray}
\label{ampl}
{\cal J}&=&{\cal J}_{PWIA}+{\cal J }_{^1S_0}
\nonumber\\
\nonumber\\
{\cal J}_{PWIA}&=& <L M_L 1 {\cal M_S}|1 M_D>
u_L ( p_0 )
Y_L^{M_L}(\widehat { p_0})
\nonumber\\
&&\Bigl\{ <{1\over 2} m_2^\prime {1\over 2} m_3|1 {\cal M_S}>
< m_1 m_2,\vec p_1,\vec p_0+\vec q/2|
t^0 -t^1
|\vec p,\vec p_0-\vec q/2, m m_2 ^ \prime >-
\nonumber\\
&&<{1\over 2} m_2^\prime {1\over 2} m_2|1 {\cal M_S}>
< m_1 m_3,\vec p_1,\vec p_0-\vec q/2|
t^0 -t^1
|\vec p,\vec p_0+\vec q/2, m m_2 ^ \prime > \}
\\
\nonumber\\
{\cal J}_{^1S_0}&=&\frac {(-1)^{1-m_2 -m_2^\prime}}{\sqrt {4\pi }} 
\delta _{m_2 ~ -m_3}
<{1\over 2} m^{\prime\prime } {1\over 2} -m_2^\prime|1 M_D> 
\\
&&\int d\vec p _0 {^\prime }
< m_1 m _2^\prime ,\vec p_1,\vec p_0^\prime +\vec q/2|
t^0 -t^1
|\vec p,\vec p_0^\prime -\vec q/2, m m ^ {\prime\prime } >
\psi _{00} ^{001} (p_0^\prime ) u_0(|\vec p_0^\prime-\vec q/2|).
\nonumber
\end{eqnarray}

\begin{figure}[h]
 \epsfysize=70mm
 \centerline{
 \epsfbox{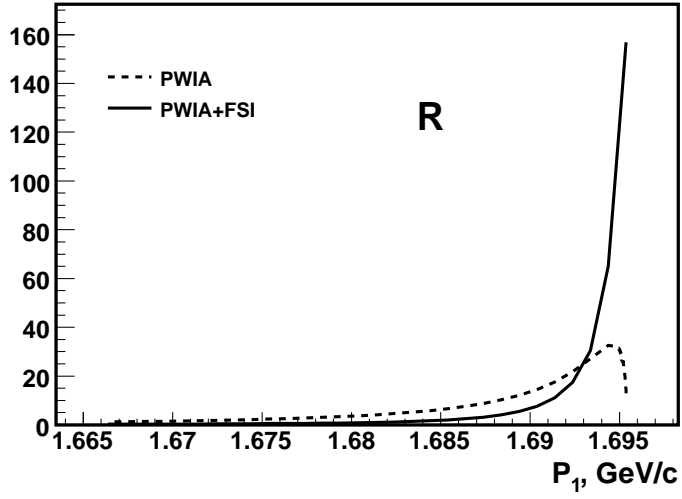}}
 \caption{$R$ ratio {\it vs.} the fast proton momentum $p_1$ at $T_n=1 ~ GeV$ }
\end{figure}

The wave function of the final
$pp$-pair  $\psi _{00} ^{001} (p_0^\prime )$ can be expressed by a series of
$\delta$-functions
\begin{equation}
\psi_{00}^{001}(p_0^\prime)=\sum_{j=1}^{N+1}C^{001}(j)\frac {\delta (p_j-p_0)}{p_j^2},
\end{equation}
where $p_j (j=1,N)$ are the grid points associated with the Gaussian nodes over $[-1,1]$ and
$p_{N+1}=p_0$ and $C(j)$ are the coefficients, which are determined from the solution of the linear algebraic 
equations system approximately equivalent to the Lippmann- Schwinger equation for two neutrons scattering
\cite {shebeko}. 

Since the $q, p_0 \ll p, p_1$ and subintegral function is suppressed at high $p_0^\prime$,
we can neglect by $q, p_0, p_0^\prime $ dependences of the high- energy $np$ t-matrix.
Then this vertex represents the free $np$ elastic scattering at angle $\theta =\pi$ and
t-matrix can be described by three independent amplitudes
\begin{eqnarray}
t_{NN}^{cm}(\theta^* =\pi)=A +
(F-B) (\vec\sigma_1 \hat q^*) (\vec\sigma_2 \hat q^*)+
B  (\vec\sigma_1 \vec\sigma_2),
\end{eqnarray}
where $\hat q^*$ is the unit vector in the beam direction.

The cross section of the $nd\to pnn$ reaction is defined by the standard manner
\begin{equation}
\sigma=(2\pi )^4~ \frac {E}{p}\cdot \frac {1}{6} \int ~ d\vec p_1 ~ d\vec p_2 ~
 \delta (M_d+E-E_1-E_2-E_3)~~|{\cal J}|^2, 
\end{equation}
where $\vec p_3=\vec p-\vec p_1-\vec p_2$ and $E_3=\sqrt {m^2+(\vec p-\vec p_1-\vec p_2)^2}$ and squared 
amplitude has the following form
\begin{equation}
|{\cal J}|^2\approx \frac {1}{2\pi}~\left( \frac {m+E}{2E}\right)^2 (2B^2+F^2)\left\{
u(p_0)+
\sum _{j=1}^{N+1} ~u(p_j)C^{001}(j)
\right\}^2
\end{equation}
We get the factorization of the squared amplitude on the two parts. 
One of them depends on the deuteron and two slow neutrons wave functions.
Other term corresponds to the spin-dependent component of the elementary
$np\to pn$ cross section.
\newpage
\begin{figure}[h]
 \epsfysize=70mm
 \centerline{
 \epsfbox{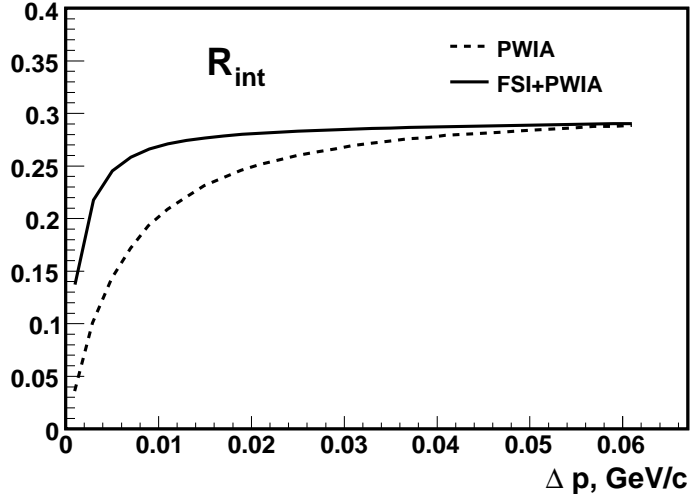}}
 \caption{Integrated R ratio as a function of $\Delta  p$ at $T_n=1 ~ GeV$ }
\end{figure}

\section{Results}
Here we consider the ratio of the $nd$ charge exchange differential cross section 
to the free $np$ scattering differential cross section.
\begin{equation}
R=\frac {d\sigma (nd\to pnn)}{dp_1 d\Omega}/\frac {d\sigma (np\to pn)}{d\Omega }
\end{equation}
 This ratio is presented
in Fig.1 as a function of the final proton momentum $p_1$.

 The solid line,which corresponds to the full calculation, has a very sharp
peak, when momentum $p_1$ is close to beam momentum $p$,
 or transfer momentum $q$ is close to zero.
This peak indicates the FSI contribution to $nd$ differential cross section.

In this region the value of the R ratio varies in 10 times,
 while transfer momentum changes on  few MeV. 
Since any experiment has the limited momentum resolution, we consider the R ratio 
integrated over $p_1$ in some region. 
\begin{equation}
 R_{int}=\int_{p-\Delta p}^{p} dp_1 R(p_1)=\int_{p-\Delta p}^{p} dp_1 \frac 
{d\sigma (nd\to pnn)}{dp_1 d\Omega}/\frac {d\sigma (np\to pn)}{d\Omega } 
\end{equation}
The integration limits change from
$p-\Delta p$ up to maximal value $p_1$ equal $p$. 
The $\Delta p$  is the difference between $p$ and $p_1$.
 The integrated R-ratio  is shown in Fig2. in dependence on the change 
 of integration  limits .

One can see, that the difference between PWIA and full calculation results is about $30 \%$ 
for $\Delta p$ equal 10 MeV,
about $15 \%$ for $\Delta p$ equal 20 MeV and these  lines are practically undistinguished, when $\Delta p$
is equal 60 MeV.

The energy dependence of the integrated R ratio is presented in  Fig.3. The integration has been performed
for $\Delta p$ equal 30 MeV. We investigate energy region only up to 1300 MeV,
 when the phase shift analysis data are exist. The dash- dotted line is 
\begin{figure}[h]
 \epsfysize=70mm
 \centerline{
 \epsfbox{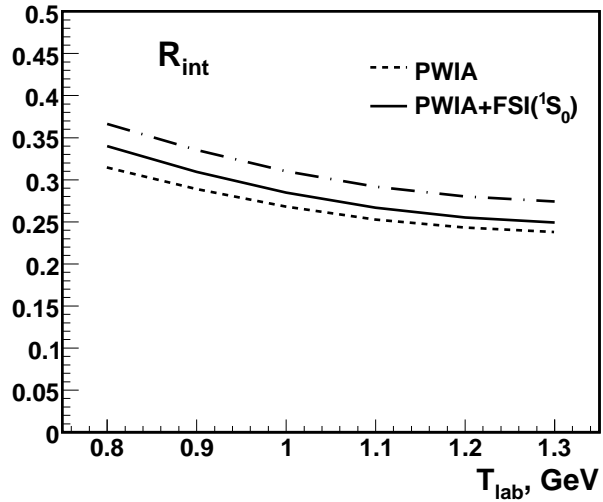}}
 \caption{Energy dependence of the integrated R-ratio}
\end{figure}
   obtained using the formula from
  ref.\cite {lbshz} with NN amplitude taken from recent phase shift analysis \cite {said}.
The difference between result obtained taking into account FSI and PWIA result
 is about 10 $\%$ for kinetic energy 800 MeV and few per cent for kinetic energy 1300 MeV.
 Thus, the contribution of the FSI decreases, when the kinetic energy is increases.

\section{Conclusion}

In this paper the $nd\to p(nn)$ reaction has been studied at the neutron kinetic energy
$T_n=0.8\div 1.3 ~ GeV$ in kinematics, when transfer momentum is close to zero.
The $R$ ratio of the $nd$ differential cross section to the
elementary $np\to pn$ one has been considered. 
 It was shown, that the final state interaction play important role, although
 the contribution of 
the FSI decreases with the increasing energy.
The factorization of the $nd$ squared amplitude has been got, what allows us to extract
the  spin dependent part of  the $np$ charge exchange  amplitude from
the $nd\to p(nn)$. 
But  obtained result will be dependent on the applied model for FSI description and
 choice of the deuteron wave function.
As a consequence, the $nd$ charge exchange reaction can not be used to 
define the precise value of the spin dependent part of the free $np$ scattering. However it is possible
to get some useful information about $np$ charge exchange process (for example, sign, approximate value etc.).

\vspace{1cm}
The author are thankful to Drs. V.P. Ladygin, F. Lehar and V.I. Sharov for fruitful discussions.
This work has been supported by the Russian Foundation for Basic Researches
under grant  $N^{\underline 0}$  04-02-17107a.

\end{document}